\documentclass[aps,twocolumn,prl,showpacs,floatfix,superscriptaddress]{revtex4}

\usepackage{amsmath,fixmath}
\usepackage{graphicx}
\usepackage{amssymb}
\usepackage{bm}

\begin{document}

\title{Nonequilibrium spin glass dynamics from picoseconds to 0.1 seconds}

\author{F.~Belletti} \affiliation{Dipartimento
  di Fisica Universit\`a di Ferrara and INFN - Sezione di Ferrara,
  Ferrara, Italy.} 

\author{M.~Cotallo} \affiliation{Instituto de Biocomputaci\'on y
  F\'{\i}sica de Sistemas Complejos (BIFI), Zaragoza, Spain.}

\author{A.~Cruz} \affiliation{Departamento
  de F\'\i{}sica Te\'orica, Universidad
  de Zaragoza, 50009 Zaragoza, Spain.} 
  \affiliation{Instituto de Biocomputaci\'on y
  F\'{\i}sica de Sistemas Complejos (BIFI), Zaragoza, Spain.}

\author{L.A.~Fernandez} \affiliation{Departamento
  de F\'\i{}sica Te\'orica I, Universidad
  Complutense, 28040 Madrid, Spain.} 
  \affiliation{Instituto de Biocomputaci\'on y
  F\'{\i}sica de Sistemas Complejos (BIFI), Zaragoza, Spain.}

\author{A.~Gordillo-Guerrero}\affiliation{Departamento de
  F\'{\i}sica, Universidad de Extremadura, 06071 Badajoz, Spain.}
\affiliation{Instituto de Biocomputaci\'on y
  F\'{\i}sica de Sistemas Complejos (BIFI), Zaragoza, Spain.} 

\author{M.~Guidetti} \affiliation{Dipartimento
  di Fisica Universit\`a di Ferrara and INFN - Sezione di Ferrara,
  Ferrara, Italy.} 

\author{A.~Maiorano} \affiliation{Dipartimento
  di Fisica Universit\`a di Ferrara and INFN - Sezione di Ferrara,
  Ferrara, Italy.}  \affiliation{Instituto de Biocomputaci\'on y
  F\'{\i}sica de Sistemas Complejos (BIFI), Zaragoza, Spain.}

\author{F.~Mantovani} \affiliation{Dipartimento
  di Fisica Universit\`a di Ferrara and INFN - Sezione di Ferrara,
  Ferrara, Italy.} 

\author{E.~Marinari} \affiliation{Dipartimento di Fisica, INFM and
  INFN, Universit\`a di Roma ``La Sapienza'', 00185 Roma, Italy.}

\author{V.~Martin-Mayor} \affiliation{Departamento de F\'\i{}sica
  Te\'orica I, Universidad Complutense, 28040 Madrid, Spain.} 
\affiliation{Instituto de Biocomputaci\'on y
  F\'{\i}sica de Sistemas Complejos (BIFI), Zaragoza, Spain.}

\author{A.~Mu\~noz Sudupe} \affiliation{Departamento
  de F\'\i{}sica Te\'orica I, Universidad
  Complutense, 28040 Madrid, Spain.} 

\author{D.~Navarro} \affiliation{Departamento  de Ingenier\'{\i}a,
  Electr\'onica y Comunicaciones and Instituto de Investigaci\'on en\\
  Ingenier\'{\i}a de Arag\'on, Universidad de Zaragoza, 50018 Zaragoza, Spain.}

\author{G.~Parisi} \affiliation{Dipartimento di Fisica, INFM and
  INFN, Universit\`a di Roma ``La Sapienza'', 00185 Roma, Italy.}

\author{S.~Perez-Gaviro} \affiliation{Instituto de Biocomputaci\'on y
  F\'{\i}sica de Sistemas Complejos (BIFI), Zaragoza, Spain.}

\author{J.~J.~Ruiz-Lorenzo} \affiliation{Departamento de
  F\'{\i}sica, Universidad de Extremadura, 06071 Badajoz, Spain.}
\affiliation{Instituto de Biocomputaci\'on y
  F\'{\i}sica de Sistemas Complejos (BIFI), Zaragoza, Spain.}

\author{S.F.~Schifano} \affiliation{Dipartimento
  di Fisica Universit\`a di Ferrara and INFN - Sezione di Ferrara,
  Ferrara, Italy.} 

\author{D.~Sciretti} \affiliation{Instituto de Biocomputaci\'on y
  F\'{\i}sica de Sistemas Complejos (BIFI), Zaragoza, Spain.}

\author{A.~Tarancon} \affiliation{Departamento
  de F\'\i{}sica Te\'orica, Universidad
  de Zaragoza, 50009 Zaragoza, Spain.} 
  \affiliation{Instituto de Biocomputaci\'on y
  F\'{\i}sica de Sistemas Complejos (BIFI), Zaragoza, Spain.}

\author{R.~Tripiccione} \affiliation{Dipartimento
  di Fisica Universit\`a di Ferrara and INFN - Sezione di Ferrara,
  Ferrara, Italy.} 
 
\author{J.L.~Velasco} \affiliation{Instituto de Biocomputaci\'on y
  F\'{\i}sica de Sistemas Complejos (BIFI), Zaragoza, Spain.}

\author{D.~Yllanes}  \affiliation{Departamento de F\'\i{}sica
  Te\'orica I, Universidad Complutense, 28040 Madrid, Spain.}
  \affiliation{Instituto de Biocomputaci\'on y
  F\'{\i}sica de Sistemas Complejos (BIFI), Zaragoza, Spain.}

\date{\today}

\begin{abstract}
  We study numerically the nonequilibrium dynamics of the Ising Spin
  Glass, for a time that spans eleven orders of magnitude, thus
  approaching the experimentally relevant scale (i.e. {\em seconds}).
  We introduce novel analysis techniques that allow to compute the
  coherence length in a model-independent way.  Besides, we present
  strong evidence for a replicon correlator and for overlap
  equivalence. The emerging picture is compatible with non-coarsening
  behavior.
\end{abstract}
\maketitle 

Spin Glasses\cite{EXPBOOK} (SG) exhibit remarkable features,
including slow dynamics and a complex space of states: their
understanding is a key problem in condensed-matter physics that enjoys
a paradigmatic status because of its many applications to glassy
behavior, optimization, biology, financial markets, social dynamics.

Experiments on Spin Glasses\cite{EXPBOOK,LETDYN1} focus on
nonequilibrium dynamics. In the simplest experimental protocol,
isothermal aging hereafter, the SG is cooled as fast as possible to
the working temperature below the critical one, $T<T_\mathrm{c}$. It
is let to equilibrate for a {\em waiting time}, $t_\mathrm{w}$. Its
properties are probed at a later time, $t+t_\mathrm{w}$. The
thermoremanent magnetization is found to be a function of
$t/t_\mathrm{w}$, for $10^{-3}<t/t_\mathrm{w}<10 $ and $t_\mathrm{w}$
in the range 50\,s\,---\,$10^4$\,s\cite{RODRIGUEZ} (see,
however,\cite{SACLAY}). This lack of any characteristic time scale is
named {\em Full-Aging}. Also the growing size of the coherent domains,
the coherence-length, $\xi$, can be measured\cite{ORBACH,BERT}. Two
features emerge: (i) the lower $T$ is, the slower the growth of
$\xi(t_\mathrm{w})$ and (ii) $\xi\sim 100$ lattice spacings, even for
$T\sim T_\mathrm{c}$ and $t_\mathrm{w}\sim 10^4$~s\cite{ORBACH}.

The sluggish dynamics arises from a thermodynamic transition at
$T_\mathrm{c}$\cite{EXPERIMENTOTC,BALLESTEROS,PALASS-CARACC}.  There
is a sustained theoretical controversy on the properties of the
(unreachable in human times) equilibrium low temperature SG phase,
which is nevertheless relevant to (basically nonequilibrium)
experiments\cite{FRANZ}. The main scenarios are the
droplets\cite{DROPLET}, replica symmetry breaking (RSB)\cite{RSB},
and the intermediate Trivial-Non-Trivial (TNT) picture\cite{TNT}.

Droplets expects two equilibrium states related by global spin
reversal. The SG order parameter, the spin overlap $q$, takes only two
values $q=\pm q_\mathrm{EA}\,.$ In the RSB scenario an infinite number
of pure states influence the dynamics\cite{RSB,FEGI,CONTUCCI-SEP}, so
that all $-q_\mathrm{EA}\!\leq\!q\!\leq\!q_\mathrm{EA}$ are reachable.
TNT\cite{TNT} describes the SG phase similarly to an antiferromagnet
with random boundary conditions: even if $q$ behaves as for RSB
systems, TNT agrees with droplets in the vanishing surface-to-volume
ratio of the largest thermally activated spin domains (i.e. the
link-overlap defined below takes a single value).

Droplets isothermal aging\cite{SUPERUNIVERSALITY} is that of a
disguised ferromagnet\footnote{Temperature chaos could spoil the
  analogy if temperature is varied during the Aging
  experiment\cite{SUPERUNIVERSALITY}.}.  A picture of isothermal aging
emerges that applies to basically all coarsening systems:
superuniversality\cite{SUPERUNIVERSALITY}.  For $T<T_\mathrm{c}\,$
the dynamics consists in the growth of compact domains (inside which
the spin overlap coherently takes one of its possible values $q=\pm
q_\mathrm{EA}$).  Time dependencies are entirely encoded in the growth
law of these domains, $\xi(t)$. The antiferromagnet analogy
suggests a similar TNT Aging behavior.

Since in the RSB scenario $q\!=\!0$ equilibrium states do exist, the
nonequilibrium dynamics starts, and remains forever, with a vanishing
order parameter.  The replicon, a critical mode analogous to magnons
in Heisenberg ferromagnets, is present for all
$T<T_\mathrm{c}$\cite{DeDominicis}. Furthermore, $q$ is not a
privileged observable (overlap equivalence\cite{FEGI}): the link
overlap displays equivalent Aging behavior.

These theories need numerics to be
quantitative\cite{RIEGER,KISKER,MPRTRL_1999,BERBOU,SUE,SUE2,SUE3,LET-NUM,LET-NSU},
but simulations are too short: one Monte Carlo Step (MCS) corresponds
to $10^{-12}$ s\cite{EXPBOOK}. The experimental scale is at $10^{14}$
MCS ($\sim 100$\,s), while typical nonequilibrium simulations reach
$\sim 10^{-5}$s. In fact, high-performance computers have been
designed for SG simulations\cite{OGIELSKI,SUE_DEF,JANUS}.

Here we present the results of a large simulation campaign performed
on the application-oriented Janus computer \cite{JANUS}. Janus allows
us to simulate the SG instantaneous quench protocol for $10^{11}$ MCS
($\sim\! 0.1$ s), enough to reach experimental times by mild
extrapolations.  Aging is investigated both as a function of time and
temperature. We obtain model-independent determinations of the SG
coherence length $\xi$. Conclusive evidence is presented for a
critical correlator associated with the replicon mode. We observe non
trivial Aging in the link correlation (a {\em nonequilibrium} test of
overlap equivalence\cite{FEGI}). We conclude that, up to experimental
scales, SG dynamics is not coarsening like.

The $D\!=\!3$ Edwards-Anderson Hamiltonian is
\begin{equation}
{\cal H}=-\sum_{\langle \boldsymbol{x}, \boldsymbol{y}\rangle } J_{\boldsymbol{x},\boldsymbol{y}} \sigma_{\boldsymbol{x}}\, \sigma_{\boldsymbol{y}}\,,\ (\langle\ldots\rangle:\mathrm{nearest\ neighbors})\,.\label{EA-H}
\end{equation}
The spins $\sigma_{\boldsymbol{x}}\!=\!\pm1$ are placed on the nodes,
$\boldsymbol{x}$, of a cubic lattice of linear size $L$ and periodic
boundary conditions. The couplings
$J_{\boldsymbol{x},\boldsymbol{y}}\!=\!\pm 1$ are chosen randomly with
$50\%$ probability, and are quenched variables. For each choice of the
couplings (one sample), we simulate two independent systems,
$\{\sigma_{\boldsymbol{x}}^{(1)}\}$ and
$\{\sigma_{\boldsymbol{x}}^{(2)}\}$. We denote by $\overline{(\cdot
  \cdot \cdot)}$ the average over the couplings.  Model (\ref{EA-H})
undergoes a SG transition at $T_\mathrm{c}=1.101(5)$\cite{PELISSETTO}.

Our $L\!=\!80$ systems evolve with a Heat-Bath
dynamics\cite{VICTORAMIT}, which is in the Universality Class of
the physical evolution.  The fully disordered starting spin
configurations are instantaneously placed at the working temperature
(96 samples at $T\!=\!0.8\!\approx\!  0.73\, T_\mathrm{c}$, 64 
at $T\!=\!0.7\!\approx\! 0.64\, T_\mathrm{c}$ and 96 at
$T=0.6\approx 0.54\, T_\mathrm{c}$).  We also perform shorter
simulations (32 samples) at $T_\mathrm{c}$, as well as $L\!=\!40$ and
$L\!=\!24$ runs to check for Finite-Size effects.

\begin{figure}[]
  \includegraphics[width=0.54\columnwidth,angle=270,trim=0 18 10 0]{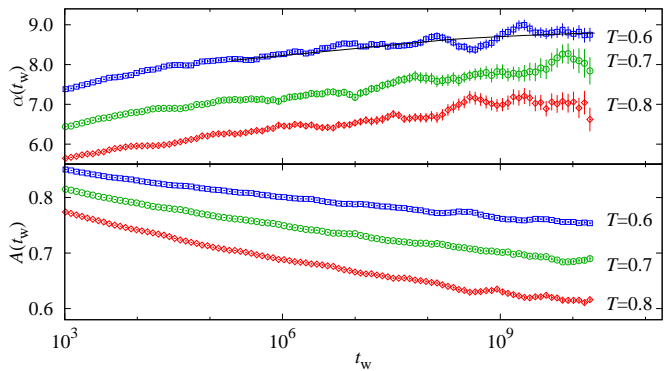}
  \caption{(Color online) Fit parameters, $A$ and $\alpha$
    ($C(t,t_\mathrm{w})=A(t_\mathrm{w})
    (1+t/t_\mathrm{w})^{-1/\alpha(t_\mathrm{w})}\,$) vs. $t_\mathrm{w}$
    for temperatures below $T_\mathrm{c}$ ($T\!=\!0.6$ line: fit (for $t_w
    >10^5$) to
    $\alpha(t_\mathrm{w})\!=\!\alpha_0+\alpha_1 \log t_\mathrm{w} +
    \alpha_2 \log^2 t_\mathrm{w}$, $\alpha_0\!=\!6.35795$, $\alpha_1\!=\!0.18605$, $\alpha_2\!=\!-0.00351835$, diagonal
    $\chi^2/\mathrm{dof}=66.26/63$). Coherent oscillations are due to
    the strong correlations of $\alpha(t_\mathrm{w})$ at neighboring
    times (neither statistical errors nor the
    fitting curve nor $\chi^2/\mathrm{dof}$ vary if one bins
    data in blocks of 5 consecutive
    $t_\mathrm{w}$).}\label{FIG-RIEGER}
\end{figure}

A crucial quantity in non equilibrium dynamics is the two-times correlation function (defined in
terms of the field 
$c_{\boldsymbol{x}}(t,t_\mathrm{w})\equiv \sigma_{\boldsymbol{x}}(t+t_\mathrm{w}) \sigma_{\boldsymbol{x}}(t_\mathrm{w})$)\cite{SUE,KISKER,RIEGER}:
\begin{equation}
C(t,t_\mathrm{w})=\overline{L^{-3}\sum_{\boldsymbol{x}} c_{\boldsymbol{x}}(t,t_\mathrm{w})}\label{CTTW}\,,
\end{equation}
linearly related to the real part of the a.c. susceptibility at
waiting time $t_\mathrm{w}$ and frequency $\omega=\pi/t$.

To check for Full-Aging\cite{RODRIGUEZ} in a systematic way, we fit
$C(t,t_\mathrm{w})$ as $A(t_\mathrm{w})
(1+t/t_\mathrm{w})^{-1/\alpha(t_\mathrm{w})}\,$ in the range
$t_\mathrm{w}\!\leq\!  t\!\leq\!10t_\mathrm{w}$ \footnote{ Because
  data at different $t$ and $t_\mathrm{w}$ are exceedingly correlated,
  for all fits in this work we consider the diagonal $\chi^2$ (i.e.
  we keep only the diagonal terms in the covariance matrix). The
  effect of time correlations is considered by first forming jackknife
  blocks\cite{VICTORAMIT} (JKB) with the data for different samples
  (JKB at different $t$ and $t_\mathrm{w}$ preserve time
  correlations), then minimizing $\chi^2$ for each JKB\cite{SUE}.  },
obtaining fair fits for all $t_\mathrm{w}\!>\! 10^3\,$.  To be
consistent with the experimental claim of Full-Aging behavior for
$10^{14}\!<\!t_\mathrm{w}\!<\!  10^{16}$\cite{RODRIGUEZ},
$\alpha(t_\mathrm{w})$ should be constant in this $t_\mathrm{w}$
range. Although $\alpha(t_\mathrm{w})$ keeps growing for our largest
times (with the large errors in\cite{SUE} it seemed constant for
$t_\mathrm{w}\!>\!10^4$), its growth slows down. The behavior at
$t_\mathrm{w}=10^{16}$ seems beyond reasonable extrapolation.

\begin{figure}[]
  \includegraphics[width=0.45\columnwidth,angle=270,trim=0 18 10 0]{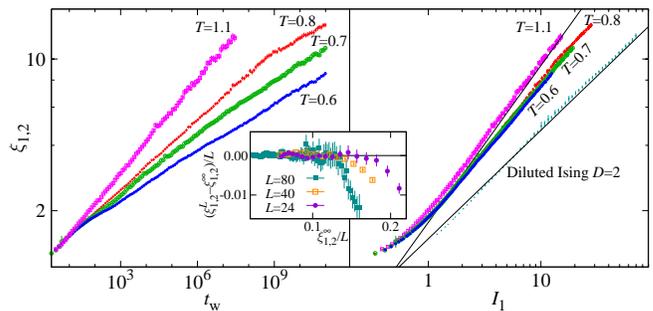}
  \caption{(Color online) {\bf Left:} SG coherence length $\xi_{1,2}$
    vs. waiting time, for $T\!\leq\!T_\mathrm{c}$.  {\bf Right:}
    $\xi_{1,2}$ vs. $I_1$, ($\xi_{1,2} \propto I_1^{1/(2-a)}$). Also
    shown data for the 2-D site-diluted Ising model ($L\!=\!4096$,
    25\% dilution, average over 20 samples, $T\!=\!0.64
    T^\mathrm{Ising}_\mathrm{c}$, $\xi_{1,2}$ and $I_1$ rescaled by 2
    for clarity).  Full lines correspond to Ising $a\!=\!0$,
    (coarsening) and to the SG,
    $a(T_\mathrm{c})\!=\!0.616$\cite{PELISSETTO}. {\bf Inset:} $[
    \xi_{1,2}^{L}(t_\mathrm{w})- \xi_{1,2}^{\infty}(t_\mathrm{w})]/L$
    vs. $\xi_{1,2}^{\infty}(t_\mathrm{w})/L$ for $T\!=\!0.8$ and
    $L\!=\!24,40$ and $80$ ($\xi_{1,2}^\infty(t_\mathrm{w})$ from
    a fit $\xi_{1,2}(t_\mathrm{w})\!=\! A(T) t_\mathrm{w}^{1/z(T)}$
    for $L\!=\! 80$ in the range $3<\xi_{1,2}<10$, see
    text).}\label{FIGXI}
\end{figure}

The coherence length is studied  from the correlations of
the replica
field $q_{\boldsymbol{x}}(t_\mathrm{w})\equiv \sigma_{\boldsymbol x}^{(1)}(t_\mathrm{w}) \sigma_{\boldsymbol{x}}^{(2)}(t_\mathrm{w})\,$,
\begin{equation}
C_4(\boldsymbol r,t_\mathrm{w})=\overline{L^{-3}\sum_{\boldsymbol x} q_{\boldsymbol x}(t_\mathrm{w}) q_{\boldsymbol x+\boldsymbol r}(t_\mathrm{w})}\,.\label{C4DEF}
\end{equation}
For $T<T_\mathrm{c}$, it is well described by\cite{MPRTRL_1999,RSB}
\begin{equation}
 C_4(\boldsymbol r,t_\mathrm{w})\sim
 r^{-a}\mathrm{e}^{-\left(r/\xi(t_\mathrm{w})\right)^b},\quad a\simeq 0.5,\ b\simeq 1.5 \,.\label{ROMANA}
\end{equation}
The actual value of
$a$ is relevant. For coarsening dynamics
$a\!=\!0$, while in a RSB scenario $a\!>\!0$ and $C_4(r,t_\mathrm{w})$
vanishes at long times for fixed $r/\xi(t_\mathrm{w})$.  At
$T_\mathrm{c}$, the latest estimate is
$a\!=\!1+\eta\!=\!0.616(9)\,$\cite{PELISSETTO}.

To study $a$ independently of a particular Ansatz as (\ref{ROMANA})
we consider the integrals
\begin{equation}
I_k(t_\mathrm{w})=\int_0^{\infty} \mathrm{d} r\, r^k C_4(r,t_\mathrm{w})\,,\label{IKDEF}
\end{equation}
(e.g. the SG susceptibility is
$\chi^\mathrm{SG}(t_\mathrm{w})\!=\!4\pi I_2(t_\mathrm{w})\,$).  As we
assume $L\!\gg\!\xi(t_\mathrm{w})$ we safely reduce the upper limit to
$L/2$.  If a scaling form $C_4(r,t_\mathrm{w})\!\sim\! r^{-a}
f(r/\xi(t_\mathrm{w}))$ is adequate at large $r$, then
$I_k(t_\mathrm{w}) \!\propto\! [\xi(t_\mathrm{w})]^{k+1-a}\,$.  It follows
that $\xi_{k,k+1}(t_\mathrm{w})\!\equiv\!
I_{k+1}(t_\mathrm{w})/I_k(t_\mathrm{w})$ is proportional to
$\xi(t_\mathrm{w})$ and $I_1(t_\mathrm{w})\!\propto\!
\xi_{k,k+1}^{2-a}\,.$ We find $\xi^{(2)}(t_\mathrm{w})\!\approx\! 0.8\,
\xi_{1,2}(t_\mathrm{w})$, where $\xi^{(2)}$ is the noisy
second-moment estimate\cite{BALLESTEROS}. Furthermore, for
$\xi_{1,2}\!>\!3$, we find $\xi_{0,1}(t_\mathrm{w})\!\approx\! 0.46\,
\xi_{1,2}(t_\mathrm{w})$, and $\xi^\mathrm{fit}(t_\mathrm{w})\!=\!1.06\,\xi_{1,2}(t_\mathrm{w})$, ($\xi^\mathrm{fit}$ from a fit to
(\ref{ROMANA}) with $a\!=\!0.4$).

Note that, when $\xi \! \ll \! L$, irrelevant distances $r\gg \xi$
largely increase statistical errors for $I_k$.  Fortunately, the very
same problem was encountered in the analysis of correlated time
series\cite{SOKAL}, and we may borrow the cure\footnote{We numerically
  integrate $C_4(r,t_\mathrm{w})$ up to a $t_\mathrm{w}$ dependent
  cutoff, chosen as the smallest integer such that
  $C_4(r^\mathrm{cutoff}(t_\mathrm{w}),t_\mathrm{w})$ was smaller than
  three times its own statistical error. We estimate the (small)
  remaining contribution, by fitting to Eq.(\ref{ROMANA}) then
  integrating numerically the fitted function from
  $r^\mathrm{cutoff}\!-\!1$ to $L/2$. Details (including
  consistency checks) will be given elsewhere\cite{INPREPARATION}.}.

The largest $t_\mathrm{w}$ where $L\!=\!80$ still represents
$L\!=\!\infty$ physics follows from Finite Size
Scaling\cite{VICTORAMIT}: for a given numerical accuracy, one should
have $L\!\geq\! k\, \xi_{1,2}(t_\mathrm{w})\,.$ To compute $k$, we
compare $\xi_{1,2}^L$ for $L\!=\!24,40$ and $80$ with
$\xi_{1,2}^\infty$ estimated with the power law described below
(Fig.~\ref{FIGXI}---inset). It is clear that the safe range is $L\ge
7\,\xi_{1,2}(t_\mathrm{w})$ at $T\!=\!0.8$ (at $T_\mathrm{c}$ the
safety bound is $L\ge 6\,\xi_{1,2}(t_\mathrm{w})$).

Our results for $\xi_{1,2}$ are shown in Fig.~\ref{FIGXI}.  Note for
$T\!=\!0.8$ the Finite-Size change of regime at
$t_\mathrm{w}\!=\!10^9$ ($\xi_{1,2}\!\sim\!  11$).  We find fair fits
to $\xi(t_\mathrm{w})\!=\! A(T) t_\mathrm{w}^{1/z(T)}$:
$z(T_\mathrm{c})\!=\!  6.86(16)$, $z(0.8)\!=\! 9.42(15)$,
$z(0.7)\!=\!11.8(2)$ and $z(0.6)\!=\!14.1(3)$, in good agreement with
previous numerical and experimental findings $z(T)\!=\!
z(T_\mathrm{c})\,T_\mathrm{c}/T$\cite{MPRTRL_1999,ORBACH}.  We
restricted the fitting range to $3\leq \xi\leq 10$, to avoid both
Finite-Size and lattice discretization effects.  Extrapolating to
experimental times ($t_\mathrm{w}\!=\!10^{14}\sim\! 100\,$s), we find
$\xi\!=\!14.0(3), 21.2(6), 37.0(14)$ and $119(9)$ for $T\!=\!0.6$,
$T\!=\!0.7$, $T\!=\!0.8$ and $T\!=\!1.1\!\approx\!T_\mathrm{c}$,
respectively, which seem fairly sensible compared with experimental
data\cite{BERT,ORBACH}.

In Fig.~\ref{FIGXI}, we also explore the scaling of $I_1$ as a
function of $\xi_{1,2}$ ($I_1\propto \xi^{2-a}$). The nonequilibrium
data for $T=1.1$ scales with $a=0.585(12)$. The deviation from the
{\em equilibrium} estimate $a=0.616(9)$\cite{PELISSETTO} is at the
limit of statistical significance (if present, it would be due to
scaling corrections). For $T=0.8, 0.7$ and 0.6, we find $a=0.442(11),
0.355(15)$ and $0.359(13)$ respectively (the residual $T$ dependence
is probably due to critical effects still felt at $T\!=\!0.8$).  Note
that ground state computations for $L\leq 14$ yielded
$a(T\!=\!0)\approx 0.4$\cite{GROUND}.  These numbers differ both from
critical and coarsening dynamics ($a\!=\!0$).

\begin{figure}[]
\includegraphics[width=0.36\columnwidth,angle=270,trim=0 10 20 0]{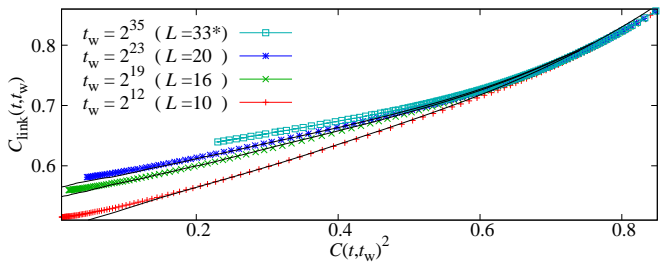}
\caption{(Color online) For appropriate $t_\mathrm{w}$ and $L$, the
  nonequilibrium $C_\mathrm{link}(t,t_\mathrm{w})$ vs. $C^2(t,t_\mathrm{w})$ at $T\!=\!0.7\,,$
  coincides with equilibrium $\left.Q_\mathrm{link}\right|_q$
  vs. $q^2$ (full lines, data from\cite{ULTRA}, see text for
  definitions).  The length-time dictionary is $L\!=\!10$ or
  $t_\mathrm{w}\!=\!2^{12}$, $L\!=\!16$ or $t_\mathrm{w}\!=\!2^{19}$ and $L\!=\!20$ or
  $t_\mathrm{w}\!=\!2^{23}$. The coherence lengths, $\xi(2^{12})\!=\!2.75(3)$,
  $\xi(2^{19})\!=\!4.23(4)$ and $\xi(2^{23})\!=\!5.40(7)$, are in the
  ratio 10:16:20. Hence, from $\xi(2^{35})$, Fig.~\ref{FIGXI}, we predict the
  equilibrium curve for $L\!=\!33\,$.}\label{FIGEQUILIBRIO}
\end{figure}

We finally address the aging properties of $C_\mathrm{link}(t,t_\mathrm{w})$
\begin{equation}
  C_\mathrm{link}(t,t_\mathrm{w})=\overline{\sum_{\langle\boldsymbol{x},\boldsymbol{y}\rangle} c_{\boldsymbol x}(t,t_\mathrm{w}) c_{\boldsymbol y}(t,t_\mathrm{w})}/(3L^3)\,.
\end{equation}
Experimentalists have yet to find a way to access $C_\mathrm{link}$,
which is complementary to $C(t,t_\mathrm{w})$ (it does not vanish if the
configurations at $t+t_\mathrm{w}$ and $t_\mathrm{w}$ differ by the spin inversion of a
compact region of half the system size).

\begin{figure}[]
\includegraphics[width=0.61\columnwidth,angle=270,trim=0 16 -10 0]{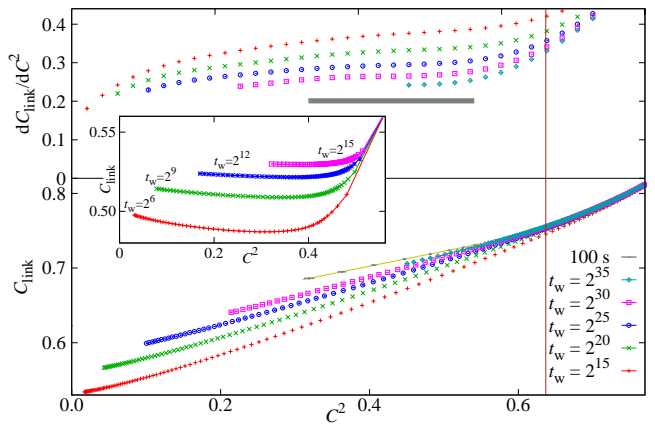}
\caption{(Color online) {\bf Bottom:}
  $C_\mathrm{link}(t,t_\mathrm{w})$ vs. $C^2(t,t_\mathrm{w})$ for
  $T=0.6$ and some of our largest $t_\mathrm{w}$ (vertical line:
  $q_\mathrm{EA}^2$ from\cite{SUE2}).  We also show our extrapolation
  of the $C_\mathrm{link}$ vs.  $C^2$ curve to $t_\mathrm{w}=10^{14}$
  ($\sim 100\,$s, see text).  {\bf Top:} Derivative of
  $C_\mathrm{link}$ with respect to $C^2$ for $T\!=\!0.6\,$.  The
  horizontal line corresponds to the slope of a linear fit of
  $t_\mathrm{w}=10^{14}$ extrapolations (the line width equals twice
  the error).  {\bf Inset:} As in bottom panel, for the ferromagnetic
  site-diluted $D\!=\!2$ Ising model (same simulation of
  Fig.~\ref{FIGXI}).  }\label{FIG-PENDIENTE}
\end{figure}

It is illuminating to eliminate $t$ as independent variable in favor
of $C^2(t,t_\mathrm{w})$, Figs.~\ref{FIGEQUILIBRIO}
and~\ref{FIG-PENDIENTE}. Our expectation for a coarsening dynamics is
that, for $C^2<q_\mathrm{EA}^2$ and large $t_\mathrm{w}$, $C_\mathrm{link}$
will be $C$-independent (the relevant system excitations are
the spin-reversal of compact droplets not affecting
$C_\mathrm{link}$). Conversely, in a RSB system new states are
continuously found as time goes by, so we expect a non constant $C^2$
dependence even if $C< q_\mathrm{EA}\,$\footnote{$C_\mathrm{link}\!\!=\!\!C^2$ in
  the full-RSB Sherrington-Kirkpatrick model.}.

General arguments tell us that the nonequilibrium $C_\mathrm{link}$
at finite {\em times} coincides with equilibrium correlation functions
for systems of finite {\em size}\cite{FRANZ},
Fig.~\ref{FIGEQUILIBRIO} ($Q_\mathrm{link}$ is just $C_4(r=1)$, while
$q$ is the spatial average of $q_{\boldsymbol x}$, Eq.(\ref{C4DEF})).  Therefore,
see caption to Fig.~\ref{FIGEQUILIBRIO}, we predict the $q^2$
dependency of the {\em equilibrium} conditional expectation
$\left.Q_\mathrm{link}\right|_q$ for lattices as large as $L=33$.

As for the shape of the curve $C_\mathrm{link}=f(C^2,t_\mathrm{w})$,
Fig.~\ref{FIG-PENDIENTE}---bottom, the $t_\mathrm{w}$ dependency is
residual.  Within our time window, $C_\mathrm{link}$ is not constant
for $C<q_\mathrm{EA}\,.$  For comparison (inset) we show
the, qualitatively different, curves for a coarsening dynamics.
Therefore, a major difference between a coarsening and a SG dynamics
is in the derivative $\mathrm{d} C_\mathrm{link}/\mathrm{d} C^2\,,$
for $C^2< q_\mathrm{EA}^2$, Fig.~\ref{FIG-PENDIENTE}---top.  We first
smooth the curves by fitting $C_{\mathrm{link}}=f(C^2)$ to the lowest
order polynomial that provides a fair fit (seventh order for
$t_\mathrm{w}\leq 2^{25}$, sixth for larger $t_\mathrm{w}$), whose
derivative was taken afterwards (Jackknife's statistical errors). 

Furthermore, we have
extrapolated both
$C_\mathrm{link}(t\!=\!r t_\mathrm{w},t_\mathrm{w})$ and $C(t\!=\!r
t_\mathrm{w},t_\mathrm{w})$ to $t_\mathrm{w}\!\approx\! 10^{14}$ ($\sim\!
100\,$s), for $r\!=\!8,4,\ldots,\frac1{16}$\footnote{For each $r$, both the link 
and the spin correlation functions are independently fitted to $a_r+b_r
  t_\mathrm{w}^{-c_r}$ (fits are stable for $t_\mathrm{w}\!>\!10^5$ with
  $c_r\approx 0.5$). These fits are then used to extrapolate the two
  correlation functions to $t_\mathrm{w}\!=\!10^{14}$.}. 
The extrapolated points for $t_\mathrm{w}\!=\!10^{14}$ fall on a straight line
whose slope is plotted in the upper panel (thick line). The
derivative is non vanishing for $C^2\!<\! q_\mathrm{EA}^2\,$, for the
experimental time scale.

In summary, Janus\cite{JANUS} halves the (logarithmic) time-gap
between simulations and nonequilibrium Spin Glass experiments.  We
analyzed the simplest temperature quench, finding numerical evidence
for a non-coarsening dynamics, at least up to experimental times (see
also\cite{LET-NSU}).  Let us highlight: {\em nonequilibrium} overlap
equivalence (Figs. \ref{FIGEQUILIBRIO},\ref{FIG-PENDIENTE});
nonequilibrium scaling functions reproducing {\em equilibrium}
conditional expectations in finite systems (Fig.~\ref{FIGEQUILIBRIO});
and a nonequilibrium replicon exponent compatible with equilibrium
computations\cite{GROUND}.  The growth of the coherence length
sensibly extrapolates to $t_w\!=\!100$ s (our analysis of dynamic
heterogeneities\cite{LET-NUM,LET-NSU} will appear
elsewhere\cite{INPREPARATION}). Exploring with Janus nonequilibrium
dynamics up to the {\em seconds} scale will allow the investigation of
many intriguing experiments.

We corresponded with  M. Hasenbusch, A. Pelissetto and E. Vicari.  Janus
was supported by EU FEDER funds (UNZA05-33-003, MEC-DGA,
Spain), and developed in collaboration with ETHlab.  We were
partially supported by MEC (Spain), through contracts 
No. FIS2006-08533, FIS2007-60977, FPA2004-02602, TEC\-2007-64188; by CAM (Spain)
and the Microsoft Prize 2007.

\end{document}